\documentclass[12pt]{article}

\usepackage{amsmath}
\usepackage{amsfonts}
\usepackage{amsthm}
\usepackage{array}
\usepackage{multirow}
\usepackage{graphicx}
\usepackage{caption}
\usepackage{subcaption}

\DeclareMathOperator{\grad}{grad}
\DeclareMathOperator{\vdiv}{div}

\newcommand{\atPoint}[1]{\big|_{#1}}

\newcommand{\dir}[1]{\partial_{#1}}

\newcommand{\LieAlgebra}[1]{\mathfrak{#1}}

\newcommand{\bvec}[1]{\mathbf{#1}}
\newcommand{\entr}{s}
\newcommand{\temp}{T}
\newcommand{\vol}{v}
\newcommand{\dens}{\rho}
\newcommand{\press}{p}
\newcommand{\energy}{\epsilon}

\newcommand{\thermVars}{\press, \temp, \energy, \vol}
\newcommand{\fullThermVars}{\thermVars,\entr}
\newcommand{\stateSurface}{L}
\newcommand{\lagrangianSurface}{\bar{L}}

\newcommand{\JetSpace}[2]{{\mathbf{J}}^{#1}#2}

\newcommand{\filsys}{\mathcal{E}}

\newtheorem{theorem}{Theorem}

\title{Non-stationary adiabatic filtration of gases in porous media}

\author{Anna Duyunova\thanks{The author is supported by the RFBR Grant No 18-29-10013.},\\
V. A. Trapeznikov Institute of Control Sciences of RAS,\\Bauman 
Moscow 
State 
Technical 
University\\
{\tt duyunova\char`_anna@mail.ru}
\and
Valentin Lychagin\thanks{The author is supported by the RFBR Grant No 18-29-10013.},\\
V. A. Trapeznikov Institute of Control Sciences of RAS,\\
University of Troms\o,\\
{\tt valentin.lychagin@uit.no}
\and
Sergey Tychkov\\
V. A. Trapeznikov Institute of Control Sciences of RAS,\\
{\tt sergey.lab06@gmail.com}}
\date{}

\begin{document}
\maketitle

\abstract{A non-stationary isentropic filtration of gases in porous media
is considered. Thermodynamics in terms of contact and symplectic geometries
is briefly discussed.
Algebra of symmetries for the PDE system is found, and the classification
of media with respect to admissible symmetries is given. Solution for one
class of media is found and the phase transitions for this solution are studied.}

{{\it Keywords}: {gas filtration, porous media, symmetry algebra, phase transitions.}}

\section{Introduction}

In this paper we consider an isentropic filtration process of gases
in a porous medium with constant porosity. Unlike \cite{LR}, where steady filtration
of gases was studied, non-stationary processes are considered here.

The system of differential equations describing
such processes consists of the following equations:

\begin{itemize}
	\item 
	the Darcy law
	\begin{equation}\label{dar}
		{\bvec{u}}=
	-\mu(\vol,\temp)\grad{\press},
	\end{equation}
	where the vector $\bvec{u}(t,x,y,z) = (u_1,u_2,u_3)$ is the gas volumetric flow,
	$\press(t,x,y,z)$ is the pressure,  $\temp(t,x,y,z)$ is the temperature,
	$\vol(t,x,y,z)$ is the specific volume of the gas, the function $\mu(\vol,\temp)$
	depends on the gas viscosity and the medium permeability;
	\item 
	the mass conservation law
	\begin{equation}\label{mass}
	q\,\vol_t + \bvec{u}\cdot \grad{\vol} = \vol \vdiv{ \bvec{u}} ,
	\end{equation}
	where constant $q$ is the medium porosity;
	\item
	 the energy conservation law, which in the case of isentropic process, has the form
	\begin{equation}\label{isentr}
	\entr_{t} + \bvec{u}\cdot\grad{\entr} =0,
	\end{equation}
	where $\entr(t,x,y,z)$ is the specific entropy.
\end{itemize}

See also \cite{Sch} for details.

The paper is organized as follows. Section \ref{sec:thermodynamics}
briefly reminds the thermodynamical principles
in terms of contact and symplectic geometries. In Section \ref{sec:symmetries},
we find point symmetries of the isentropic filtration PDE system. The classification
of porous media with respect to admissible symmetries groups is given.
In Section~\ref{sec:solutions}, we find a solution that is invariant
with respect to a certain symmetries subgroup. This leads to an ODE system,
which can be solved explicitly. We find solution for the
ideal gas model and a certain class of media.
Then the ideal gas solution is used
to construct a solution for real gases described by van der Waals
equations. Finally, possible phase transitions are studied.

Many of the computations in this paper were done in Maple with the
Differential Geometry package by I.~Anderson and his team. Maple files with
the most important computations in this paper can be found on the web-site
{\it http://d-omega.org}.

\section{Thermodynamics}\label{sec:thermodynamics}

Here we briefly recall the thermodynamical principles
expressed in terms of contact and symplectic geometries
that we need for the further discussion
(for details see~\cite{L}, \cite{LR}, \cite{Duyunova2017}). 

Consider a 5-dimensional contact manifold $\mathbb{R}^5$
equipped with the coordinates $(\fullThermVars)$ and the contact 1-form 
\[
\theta = d\entr-\temp^{-1} d\energy-{\temp}^{-1}\press d \vol,
\]
here $\energy$ is the specific energy, and $\vol=\dens^{-1}$ is the specific volume.

The thermodynamical states are a 2-dimensional Legendrian manifold
$\stateSurface$, i.e.~such surface $\stateSurface$
that the first law of thermodynamics $\theta\atPoint{\stateSurface}=0$ holds.

Using the projection $\pi : \mathbb{R}^5 \rightarrow \mathbb{R}^4$, 
$\pi : \left( \fullThermVars \right) \longmapsto
\left( \thermVars  \right)$,  eliminate the specific entropy $\entr$
from the description of the thermodynamic states. The restriction of this
projection on the state surface $\stateSurface$ leads to a Lagrangian
manifold $\lagrangianSurface$ in the 4-dimensional symplectic
space $\mathbb{R}^4$ equipped with the structure form 
\[
\Omega =-d\theta = \temp^{-1} d\press\wedge d\vol-  
 {\temp}^{-2} d\temp\wedge (d\energy  + \press d\vol) .
\]

Therefore,  the thermodynamic states
can be considered as the Lagrangian submanifolds
in the symplectic space $(\mathbb{R}^4, \Omega)$, and it can be defined by the equations 
\begin{equation}\label{eq:Therm}
\left\lbrace
\begin{aligned}
&f(\thermVars )=0,\\
&g(\thermVars )=0
\end{aligned}
\right.
\end{equation}
if  
\begin{equation}\label{eq:lagr}
[f,g]=0 \, \text{ on } \, \lagrangianSurface,
\end{equation}
where $[f,g]$ is the Poisson bracket with respect to the symplectic form $\Omega$.

In order to find the state manifolds $\lagrangianSurface$
for real gases, we consider the following two equations:
\[
\left\lbrace
\begin{aligned}
&f(\thermVars )=\press-A(\vol,\temp),\\
&g(\thermVars )=\energy -B(\vol, \temp).
\end{aligned}
\right.
\]

The first equation is called thermic equation of state, and the second one is
called caloric equation of state. Then the compatibility condition \eqref{eq:lagr} for them has the form
\[
(\temp^{-2}B)_{\vol}=(\temp^{-1}A)_{\temp}.
\]
 
Moreover, the following theorem is valid.

\begin{theorem}
	Thermodynamical states of real gases are defined by Massieu-Planck
    potential function $\phi(\vol,\temp)$ and have the following form:
    \begin{equation}\label{therm}
    \press = R \temp \phi_{\vol}, \quad
    \energy = R \temp^2 \phi_{\temp}, \quad
    \entr = R(\phi + \temp \phi_{\temp}),
    \end{equation}
    where function $\phi$ has the following expression in terms of virial coefficients $A_k$,
    \begin{equation}
    \phi(\vol,\temp) =\frac{n}{2}\ln{\temp} + \ln{\vol} - A_1(\temp)\vol^{-1} - \frac{1}{2}A_2(\temp)\vol^{-2} - \ldots - \frac{1}{k}A_k(\temp)\vol^{-k} -\ldots. 
    \end{equation}
    The domain of applicable states on the plane $(\vol,\temp)$ is given by inequalities
    \begin{equation}
    \phi_{\vol\vol}< 0, \quad \temp\phi_{\temp\temp}+2\phi_{\temp}>0.
    \end{equation}
    Phase transitions occur near the curve
    \[\phi_{\vol\vol}=0.\]
\end{theorem}

Thus, by the system $\filsys$ of differential equations describing the isentropic filtration process of gases  we mean the differential equations \eqref{dar}---\eqref{isentr} and the equations of state~\eqref{therm}.


\section{Symmetry Lie algebra}\label{sec:symmetries}

By a symmetry of the system $\filsys$ we mean a point symmetry, i.e.
a vector field $X$ on the $0$-jet space such that its second prolongation
$X^{(2)}$ is tangent to the submanifold $\filsys^{(2)}\subset \JetSpace{2}{(4,7)}$.

Using the standard techniques of the symmetries computations we obtain
(see the Maple file) the following result.

\begin{theorem}
	The Lie algebra $\LieAlgebra{g}$ of point symmetries of
	the system $\filsys$ of differential equations describing the isentropic
	filtration process of real gases in an arbitrary porous medium
	is generated by the vector fields 
	\begin{equation} 
	\begin{aligned} 
	&X_1=\dir{x},\qquad
	X_5=y\,\dir{x}-x\,\dir{y}, \phantom{hjkhkjhkhgoo.} \quad    \\ 
	&X_2=\dir{y},\qquad 
	X_6=z\,\dir{x}-x\,\dir{z}, \quad \\
	&X_3=\dir{z},\qquad 
	X_7=z \,\dir{y}-y\,\dir{z},  \quad \\
	&X_4=\dir{t},\qquad X_{8}=2t\,\dir{t}+x\,\dir{x}+y\,\dir{y}+z\,\dir{z}.
	\end{aligned} 
	\end{equation}
\end{theorem}

So, transformations corresponding to elements of the algebra  $\LieAlgebra{g}$ are compositions of the translations, the rotations  ${SO}(3)$  and the scale  transformation $X_8$. 

Consider the case when the gas satisfies ideal gas
model, in other words, the thermodynamic states
are given by the potential function 
\[
\phi(\vol,\temp) =\frac{n}{2}\ln{\temp} + \ln{\vol} .
\]

Denote the corresponding system of differential equations $\filsys_{id}$.

Depending on the properties of the gas and the rigid medium, we use
different functions $\mu(\vol,\temp)$ and, accordingly,
the algebra of point symmetries has different
additional symmetries.  

\vspace{12pt}
\begingroup
\setlength{\tabcolsep}{10pt}
\renewcommand{\arraystretch}{2.2}
\begin{tabular}{||b{0.3\linewidth}|b{0.4\linewidth}||}
	\hline
	$\mu(\vol,\temp) = f(\vol)\temp^\alpha$ &
	$X_{9}=(1+\alpha)\,t\,\dir{t} - \temp\,\dir{\temp}$ \\
	\hline
	$\mu(\vol,\temp) = f(\temp) \vol^{\alpha}$ & 
	$X_{9}=(1-\alpha)\,t\,\dir{t} + \vol\,\dir{\vol}$ \\
	\hline
	\multirow{2}{*}{$\mu(\vol,\temp) = \alpha{\vol}^{\beta} T^{\gamma}$}
	 &$X_{9}=t\,\dir{t} - \dfrac{1}{\beta+\gamma}(\vol\,\dir{\vol}+\temp\,\dir{\temp})$ \\
	&$X_{10}= (1+\gamma) \vol \,\dir{\vol} + (1-\beta)\temp
	\,\dir{\temp}$\\
	\hline
	$\mu(\vol,\temp) = \alpha \left( \dfrac{\temp}{\vol} \right)^{\beta}$ &
	$X_{9}=(1+\beta)\,t\,\dir{t} +  \vol \,\dir{\vol}$\\
	\hline
	$\mu(\vol,\temp) = f(\vol \temp) \vol^{3-q}$ &
	$X_{9}=(q-1)\,t\,\dir{t} + \vol\,\dir{\vol} - \temp\,\dir{\temp}$\\
	\hline
\end{tabular}
\vspace{12pt}
\endgroup

Summarizing, we get the following result.

\begin{theorem}
		The Lie algebra $\LieAlgebra{g_{id}}$ of point symmetries of the  system $\filsys_{id}$ of differential equations describing the isentropic filtration of ideal gases is generated by the vector fields 
		$ X_1, X_2, \ldots, X_8 $ and by one or two additional symmetries which form depends on the particular properties of a medium, i.e. depends on the function $\mu(\vol,\temp)$.   
\end{theorem}

\section{Invariant solutions}\label{sec:solutions}

In this section we find some invariant solutions of the system $\filsys$. 

In order to find these solutions consider the subalgebra 
\[
\mathfrak{g_4}= \mathfrak{so}(3)\oplus \langle X_8\rangle \subset\mathfrak{g}.
\]
The corresponding Lie group has three-dimensional orbits and 
the invariant of its action has the form
\[
r^2=\dfrac{x^2+y^2+z^2}{t}.
\]

We find a solution that is invariant with respect to the
action of this group.

\subsection{Ideal gas solution}

First, we consider the case of ideal gas.  
Then reduction of system $\filsys_{id}$
with respect to the action of $\mathfrak{g_4}$ leads to
the system of ODEs
\[
\left\lbrace 
\begin{aligned} 
&\left(2R\mu(\vol,\temp) \left(\vol_{r}\temp-\vol\temp_{r} \right) - r\vol^2    \right) 
\left(2\vol_{r}\temp+n\vol\temp_{r} \right) =0 ,\\
&2R\mu(\vol,\temp) \left( r\vol \left(\vol_{rr}\temp-\vol\temp_{rr} \right) 
+ \left(3r\vol_{r}-2\vol \right)\left(\vol_{r}\temp-\vol\temp_{r} \right)    \right)+\\ 
&\phantom{WWWWWWwwwwwwwwwwwww} r\vol \left(2R\mu_r(\vol,\temp) \left(\vol_{r}\temp-\vol\temp_{r} \right) + 
qr\vol\vol_{r}
\right) =0 . 
\end{aligned}
\right. 
\]

Note that the second factor in the first equation corresponds the case when the entropy $\entr$ is constant.

The proof of the following theorem can be found in the Maple file.

\begin{theorem}\label{th:igsol}
The $\mathfrak{g_{4}}$-invariant solution of the system $\filsys_{id}$
for the case of ideal gas depends on the properties of a porous medium and has the form
\begin{align*}
&\vol(r) = R C_1  r^{\frac{3}{1-q}},\\
&\press(r) = -\frac{1}{2} \int  \frac{r}{\mu(\vol, \temp)} \, dr,\\
&\temp(r)=\frac{\press(r)\vol(r)}{R},
\end{align*}
where $r=\sqrt{\dfrac{x^2+y^2+z^2}{t}}$, $t>0$ and $C\in\mathbb{R}$.
\end{theorem}

Consider some cases of the function $\mu(\vol, \temp)$ from the table above and corresponding expression for the pressure.
\begin{enumerate}
	\item $\mu(\vol,\temp) = \alpha \left( \dfrac{\temp}{\vol} \right)^{\beta}$,
	$\beta\neq-1$,\qquad
	$\press(r)=R\left(\dfrac{(1+\beta)(C_2-r^2)}{4\alpha R}\right)^
	{\frac{1}{1+\beta}}$
	\item $\mu(\vol,\temp) = \alpha \dfrac{\vol}{\temp}$,\qquad
	$\press(r)=C_2\exp\left(-\dfrac{r^2}{4\alpha R}\right)$
	\item $\mu(\vol,\temp) = \alpha{\vol}^{\beta} T^{\gamma}$,\quad
	$ \press(r)=\left(C_2 + \dfrac{C_1^{-\beta-\gamma}(1+\gamma)(1-q)}
	{2\alpha R^{\beta}(3\gamma+3\beta+2q-2)}r^{\frac{3\gamma+3\beta+2q-2}{q-1}} \right)^{\frac{1}{1+\gamma}}$
	\item $\mu(\vol,\temp)=\alpha\vol^{\beta}\temp^{-1}$,\quad
	$\press(r)=C_2\exp\left(\dfrac{C_1^{1-\beta}(1-q)}{2\alpha
	R^{\beta}(3\beta+2q-5)}r^{\frac{3\beta+2q-5}{q-1}}\right)$
\end{enumerate}

\textbf{Example.}
Let us write a solution for a certain gas and medium for
the second case, where $\mu(\vol,\temp)=\alpha\frac{\vol}{\temp}$.
This gives us understanding when the solution is applicable and has
physical sense.

For example, we consider methane as the gas, and the values
of parameters are the following $q=0,55$, $\alpha \approx 5\cdot 10^{-4}$, $C_1 \approx 2,7\cdot 10^{-3}$, $C_2 \approx 3 \cdot 10^{5}$.  

Instead of a domain in terms of the invariant $r$, we present
it as a region on the plane of distance $d=\sqrt{x^2+y^2+z^2}$ (vertical axis)
and time $t$ (horizontal axis).

\begin{figure}[h]
	\centering
	\begin{subfigure}[b]{0.45\textwidth}
		\includegraphics[width=\textwidth]{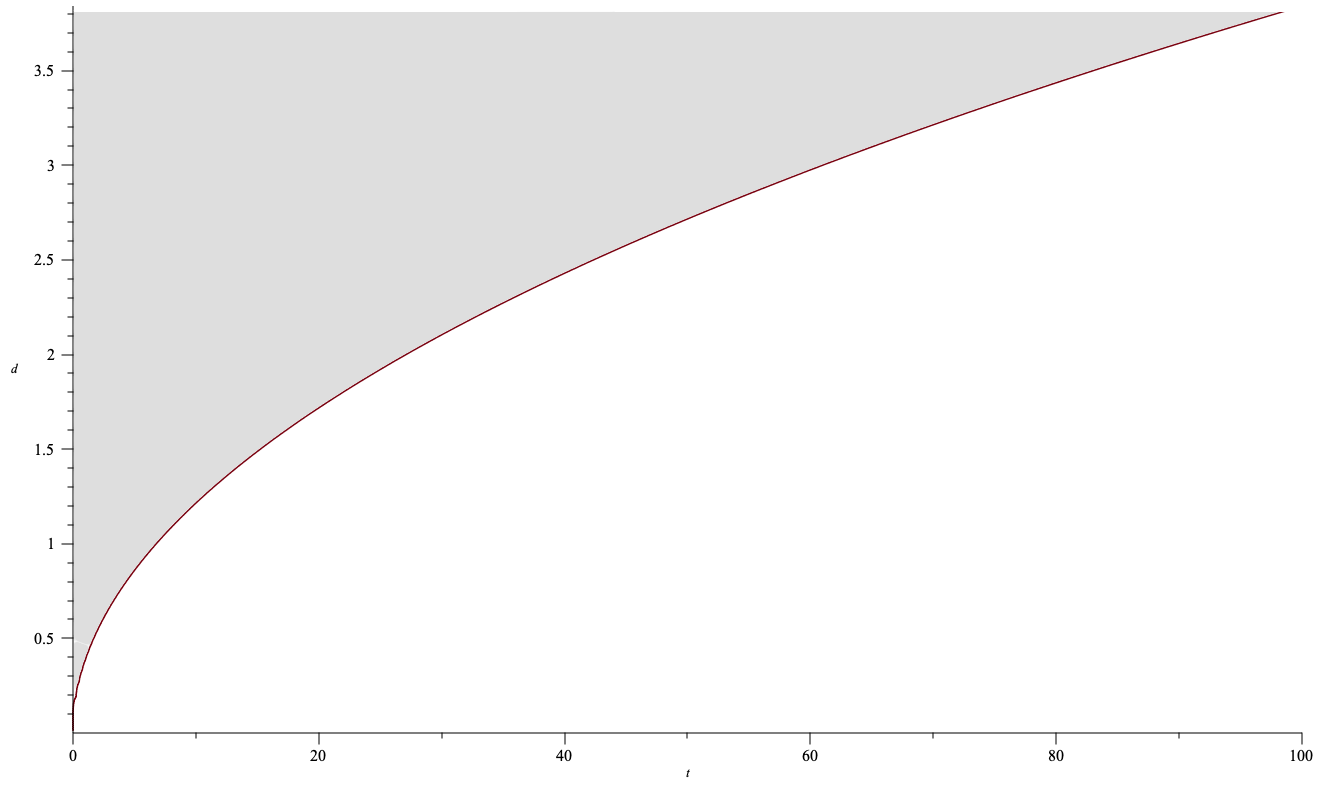}
		\caption{}\label{fig:dens}
	\end{subfigure}  \qquad
	\begin{subfigure}[b]{0.45\textwidth}
		\centering
		\includegraphics[width=\textwidth]{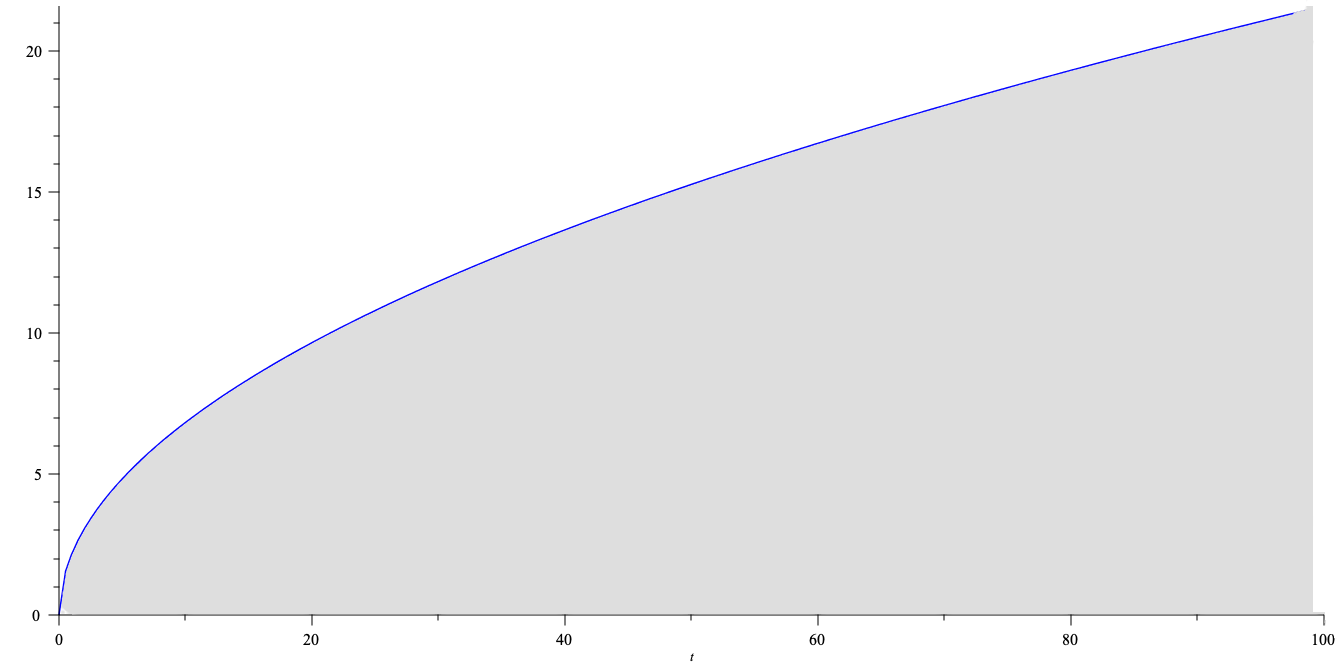}
		\caption{}\label{fig:press}
	\end{subfigure}
\begin{subfigure}[b]{0.45\textwidth}
	\centering
	\includegraphics[width=\textwidth]{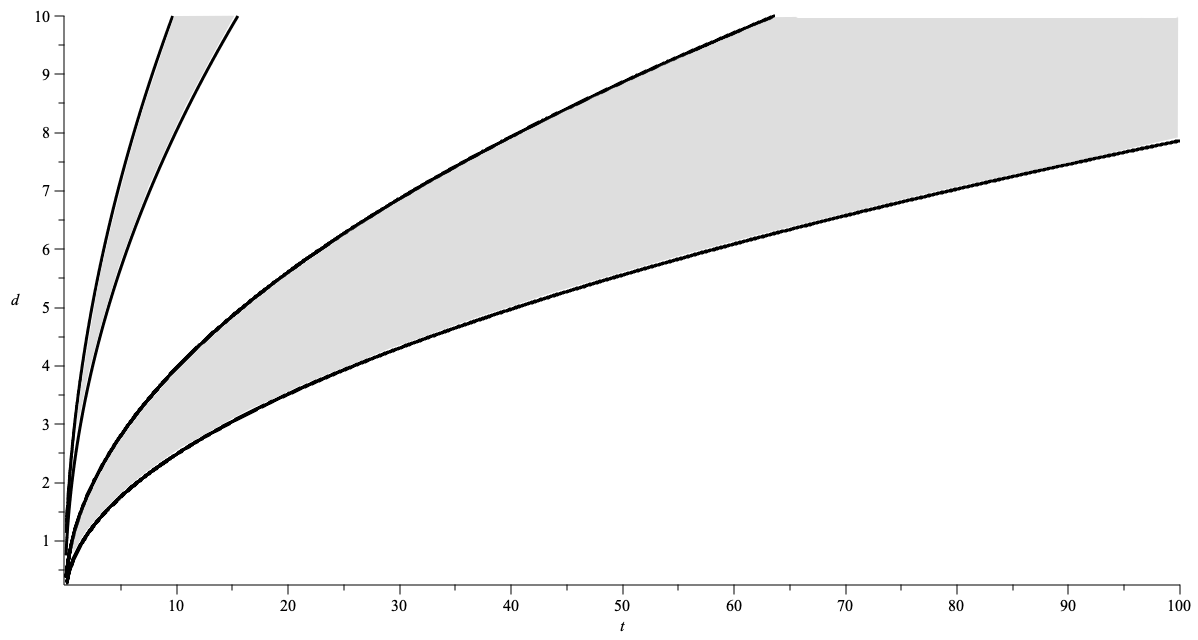}
	\caption{}\label{fig:temp}
\end{subfigure}  \qquad
\end{figure}

In the Figure \ref{fig:dens} the gray-filled region
shows where the gas density is sufficiently small to fit into
the ideal gas model. Since the maximum pressure equals $C_2$, it can be
set small enough. But the region with negligibly small pressure should
also be excluded as physically impossible.
The Figure \ref{fig:press} depicts this region. The similar
restrictions are imposed on the gas temperature. The grey-filled region
in the Figure \ref{fig:temp} is where the solution for temperature
is in certain bounds, for example between the melting and kindling points.

\begin{figure}[h]
	\centering
	\includegraphics[width=0.8\textwidth]{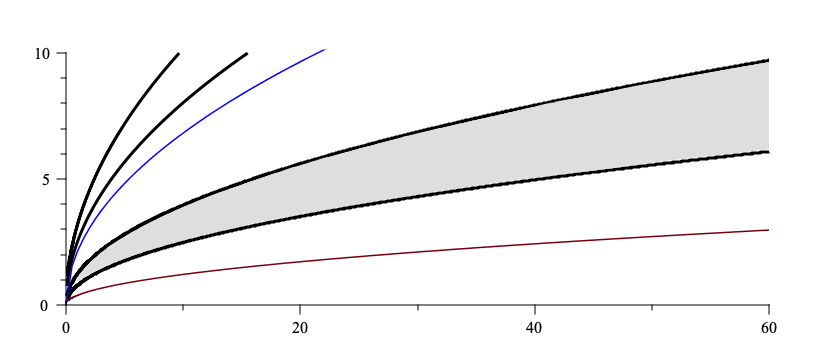}
	\caption{}\label{fig4}
\end{figure}

The intersection of all three regions is presented
in the Figure \ref{fig4}.

\subsection{van der Waals gas}

In this section we find an asymptotic solution of the system $\filsys$
for a real gas model. We use asymptotic in terms of virial coefficients and, for example, consider the model of van der Waals gas. 


Recall that thermodynamic state of a van der Waals is given by
the Massieu-Planck potential of the form:
\[
\phi=\frac{n}{2}\ln\temp+\ln(\vol-b)+\frac{a}{R\vol\temp}=
\frac{n}{2}\ln\temp+\ln\vol-\dfrac{b}{\vol}-\frac{a}{R\vol\temp}
+\mathrm{o_2},
\]
where $a, b$ are the gas parameters, and $\mathrm{o_2}$ is the terms of the second order of $a$ and $b$. Denote by $\filsys_{w}$  the corresponding system of differential equations. 

Note the value of function $\phi$, if $a=b=0$,
corresponds to the ideal gas potential, so it is a zero order approximation for real gases.

Further we look for an asymptotic solution for $\filsys_w$ in the form
\begin{equation} \label{as1}
\left\lbrace
\begin{aligned}
& \temp = \temp_0 + a\temp_1 + b\temp_2 + \mathrm{o_2},\\
& \press = \press_0 + a\press_1 + b\press_2 + \mathrm{o_2},\\
& \vol = \vol_0 + a\vol_1 + b\vol_2 + \mathrm{o_2}, \\
& \entr = \entr_0 + a\entr_1 + b\entr_2 + \mathrm{o_2}, \\
& \bvec{u} = \bvec{u}_0 + a\bvec{u}_1 + b\bvec{u}_2 + \mathrm{o_2},  
\end{aligned}
\right.
\end{equation}
where functions $\temp_0$, $\press_0$, $\vol_0$, $\entr_0$, and $\bvec{u}_0$
are the solution of the system $\filsys_{id}$,
and $\temp_1$, $\temp_2$, $\press_1$, $\dots$, $\bvec{u}_2$ are
the first-order corrections.

For this we write function $\mu$ in the form
\[
\mu(\vol, \temp)
= \mu(\vol_0, \temp_0) + \mu_{\vol}(\vol_0,\temp_0)
\left(a\vol_1 + b\vol_2\right) + \mu_{\temp}(\vol_0,\temp_0)
\left(a\temp_1 + b\temp_2\right)
+\mathrm{o_2}
\]
and substitute this expression and \eqref{as1} into the system $\filsys$. Collecting coefficients for $a$ and $b$ we get the following system
\[
\left\lbrace
\begin{aligned}
&\bvec{u}_1 + \mu(\vol_0,\temp_0)\grad{\press_1} + \grad{\press_0} \left( \vol_1 \mu_{\vol}(\vol_0,\temp_0) + \temp_1 \mu_{\temp}(\vol_0,\temp_0) \right) =0 , \\
& \bvec{u}_2 + \mu(\vol_0,\temp_0)\grad{\press_2} + \grad{\press_0} \left( \vol_2 \mu_{\vol}(\vol_0,\temp_0) + \temp_2 \mu_{\temp}(\vol_0,\temp_0) \right) 
	 = 0,\\
& q (\vol_1)_{t} + \bvec{u}_1\cdot \grad{\vol_0} +\bvec{u}_0\cdot \grad{\vol_1} - \vol_0 \vdiv{ \bvec{u}_1} - \vol_1 \vdiv{ \bvec{u}_0}   =0 , \\ 
& q (\vol_2)_{t} + \bvec{u}_2\cdot \grad{\vol_0} +\bvec{u}_0\cdot \grad{\vol_2} - \vol_0 \vdiv{ \bvec{u}_2} - \vol_2 \vdiv{ \bvec{u}_0}  =0 , \\ 
& (\entr_1)_{t} + \bvec{u}_0\cdot\grad{\entr_1} + \bvec{u}_1\cdot\grad{\entr_0} =0, \\
& (\entr_2)_{t} + \bvec{u}_0\cdot\grad{\entr_2} + \bvec{u}_2\cdot\grad{\entr_0} =0.
\end{aligned}
\right.
\]

So in order to find the first order corrections, we need to solve the linear system of ODEs on the functions $\temp_1$, $\temp_2$, $\press_1$ $\dots$, $\bvec{u}_2$.

For example, consider the invariant solution for the function 
\[
\mu(\vol,\temp) = \alpha \dfrac{\vol}{\temp}.
\]

As in the previous section, we are looking for a $\mathfrak{g}_4$-invariant
solution, i.e. the first-order corrections $\temp_1$, $\temp_2$, $\press_1$ $\dots$, $\bvec{u}_2$ also depend on the invariant $r$.

Substituting the expression for the function 
$\mu(\vol,\temp)$ and the solution of the system $\filsys_{id}$ into the our system  we get a linear system of ODEs that can be found in the Maple file. 
Solution of this system delivers the first-order corrections for volume and temperature
\[
\vol_1(r)=0, \quad \vol_2(r)=0,
\]
\[
\temp_1(r) =\left( \dfrac{6}{C_1 R^2 (q-1)}
  \int \exp\left(\frac{r^2}{4\alpha R}\right) r^{\frac{q-7}{1-q}} \, dr
+ C_3 \right) r^{\frac{3}{1-q}} \exp\left(-\dfrac{r^2}{4\alpha R}\right),
\]
\[
\temp_2(r) = \left(   \frac{C_2(q-1)r^2}{2(2q+1)\alpha R^2}
- \frac{C_2}{R}
  +C_4 r^{\frac{3}{1-q}}\right)
\exp\left(-\dfrac{r^2}{4\alpha R}\right).
\]

\begin{figure}[h]
	\centering
	\begin{subfigure}[b]{0.40\textwidth}
		\includegraphics[width=\textwidth]{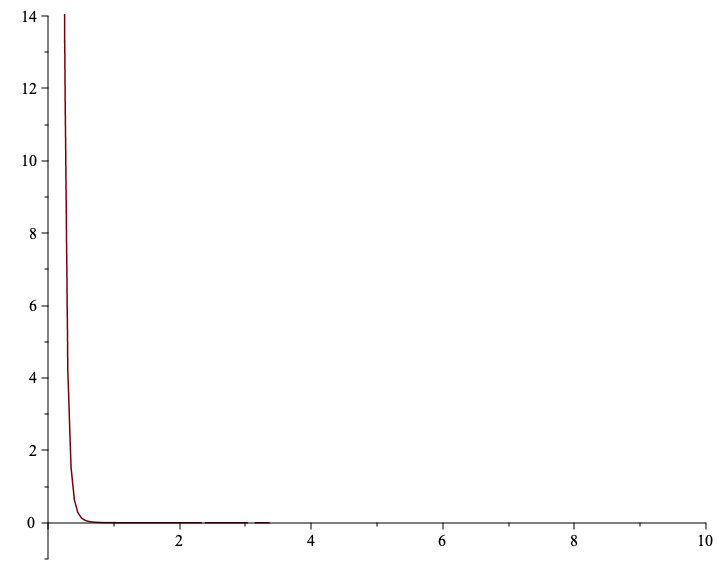}
		\caption{}\label{fig:temp1}
	\end{subfigure}  \qquad
	\begin{subfigure}[b]{0.40\textwidth}
		\centering
		\includegraphics[width=\textwidth]{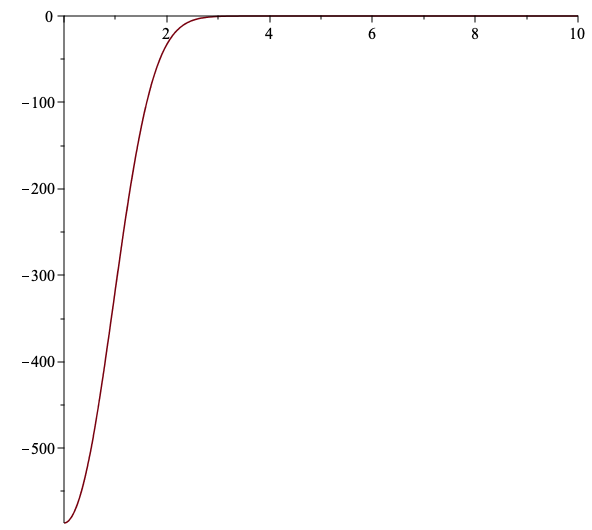}
		\caption{}\label{fig:temp2}
	\end{subfigure}
\end{figure}

\textbf{Example.} Let us draw plots for the temperature
first-order corrections
for the example we considered above. Given values for constants
$a\approx 9\cdot 10^{-5}$, $b\approx 3\cdot 10^{-3}$, $C_3=0$, $C_4=0$
the graphics of the functions $T_1$ and $T_2$ are presented on the
Figures \ref{fig:temp1}, \ref{fig:temp2}
correspondingly.

The last question we investigate is whether phase transitions occur in the
filtration process described by the solution.

\begin{figure}[h]
	\centering
	\includegraphics[width=0.55\textwidth]{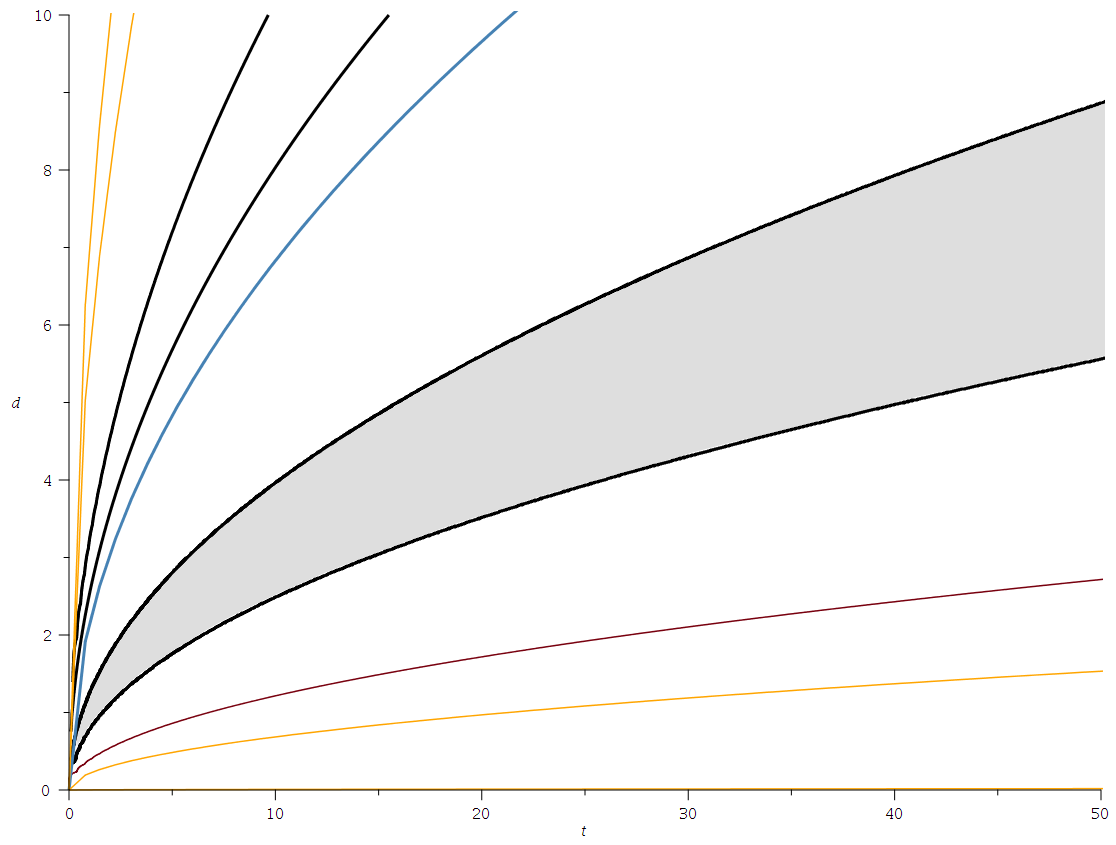}
	\caption{}\label{fig5}
\end{figure}

In terms of the Massieu-Planck potential the phase coexistence curve
on the plane $(\vol, \temp)$ is given by equations \cite{L}
\[
\left\lbrace
\begin{aligned}
&\phi_{\vol}(\vol_1, \temp)-\phi_{\vol}(\vol_2, \temp)=0,\\
&\phi(\vol_2, \temp)-\phi(\vol_1, \temp)+
\vol_1\phi_{\vol}(\vol_1, \temp)-\vol_2\phi_{\vol}(\vol_2, \temp)=0,
\end{aligned}
\right.
\]
where $\vol_1$ and $\vol_2$ are the specific volumes of phases.
Substituting the solution into these equations we
obtain four phase transition curves on the plane $(d,t)$, see the orange curves
in the Figure \ref{fig5} . Note that, though, one of these curves is practically
indistinguishable from the $t$-axis it is also a parabola.

We see that for this solution phase transitions do not occur in
the domain when it has physical sense.


\end{document}